\documentclass[final,5p,times]{elsarticle}

\journal{Computational and Structural Biotechnology Journal}

\usepackage{amsmath,amssymb}
\usepackage{graphicx}
\usepackage{dcolumn}
\usepackage{bm}
\usepackage{cancel}
\usepackage{color}
\usepackage{float}
\usepackage{xcolor}

\begin{document}

\begin{frontmatter}

\title{Holographic nature of critical quantum states of proteins}

\author{Eszter Papp\corref{cor1}}
\author{Gábor Vattay\corref{cor1}}
\cortext[cor1]{Corresponding author. Email: \texttt{gabor.vattay@ttk.elte.hu}}
\address{Department of Physics of Complex Systems, Institute for Physics and Astronomy,\\
Eötvös Loránd University, H-1053 Budapest, Egyetem tér 1-3., Hungary}

\begin{abstract}
The Anderson metal-insulator transition is a fundamental phenomenon in condensed matter physics, describing the transition from a conducting (metallic) to a non-conducting (insulating) state driven by disorder in a material. At the critical point of the Anderson transition, wave functions exhibit multifractal behavior, and energy levels display a universal distribution, indicating non-trivial correlations in the eigenstates. Recent studies have shown that proteins, traditionally considered insulators, exhibit much higher conductivity than previously assumed. In this paper, we investigate several proteins known for their efficient electron transport properties. We compare their energy level statistics, eigenfunction correlation, and electron return probability to those expected in metallic, insulating, or critical states. Remarkably, these proteins exhibit properties of critically disordered metals in their natural state without any parameter adjustment. Their composition and geometry are self-organized into the critical state of the Anderson transition, and their fractal properties are universal and unique among critical systems. { Our findings suggest that proteins' wave functions may fulfill ``holographic'' area laws, since their correlation fractal dimension is  \(d_2\approx 2\).}
\end{abstract}


\end{frontmatter}


The Anderson metal-insulator transition\cite{evers2008anderson} (MIT) is a fundamental phenomenon in condensed matter physics, describing the transition from a conducting (metallic) to a non-conducting (insulating) state driven by disorder in a material. As proposed by P.W. Anderson, this transition occurs due to the localization of electronic wave functions induced by random impurities or lattice defects. In a metallic phase, electrons can travel through the material as their wave functions are extended and correlated; random matrix theory\cite{altshuler1988repulsion,guhr1998random} describes the statistics of energy levels. However, as the disorder increases, these wave functions become increasingly localized, leading to an insulating state where eigenstates are spatially localized and uncorrelated; the energy levels follow Poisson statistics.

The critical point of the Anderson transition is characterized by a unique set of properties where wave functions exhibit multifractal\cite{castellani1986multifractal,pook1991multifractality} behavior. {At the critical point, the level statistics deviate from the random matrix and Poisson distributions. Instead, they display a universal distribution\cite{shklovskii1993statistics} that is intermediate between the metallic and insulating phases, highlighting the presence of non-trivial correlations in the eigenstates\cite{chalker1990scaling,huckestein1994relation}.} The critical state has some peculiar features. For example, it has been argued\cite{feigel2010fractal,burmistrov2012enhancement} that if one could artificially create a critically disordered metal in which the level of disorder is nearly perfectly adjusted to the critical point of the Anderson transition, a high temperature {\em fractal superconducting state} can arise due to the power-law dependence of the superconductor-insulator transition temperature on the interaction constant of the electron-electron attraction. It has also been found\cite{ng2006critical,vattay2014quantum} that systems near the critical point of MIT can have long coherence times and coherent transport at the same time, and some biological molecules prefer this state\cite{vattay2015quantum}.  

In the last ten years, measuring the conductivity of small-scale protein samples has become feasible. The previous belief that proteins are good insulators has been proven wrong; they exhibit much higher conductivity than previously assumed\cite{amdursky2014electronic,zhang2017observation,bostick2018protein}. Protein conductance can occur over long distances\cite{zhang2019electronic} and in a temperature-independent manner\cite{kayser2019solid} from four Kelvins to ambient temperatures, which is atypical for most conductive materials. In this study, we investigate several proteins known for efficient electron transport properties. We compare their energy level statistics, eigenfunction correlation, and electron return probability to those expected in metallic, insulating, or critical states. Remarkably, as we will show, these proteins exhibit properties of {\em critically disordered metals} without any parameter adjustment in their natural state. Their composition and geometry are self-organized into the critical state of the Anderson transition. { Moreover, their fractal properties appear to be universal and unique among critical systems, as their numerically determined correlation fractal dimension \(d_2\) is approximately 2, which would suggest that their wave functions fulfill ``holographic'' area laws.}

{Our investigation includes several proteins\cite{proteins} that exhibited efficient electron transport properties.} A recent study on {\em Bacteriorhodopsin} demonstrated exceptional long-range charge transport and low activation energy, challenging common electron transport models due to unusual protein-electrode energetics and transport lengths\cite{bera2023near}. Similarly, {\em Cytochrome C}, essential for electron transfer, exhibited temperature-independent conductance between 30 and $\sim130\,$K, suggesting tunneling by superexchange as a transport mechanism~\cite{amdursky2014solid}. {\em Met-Myoglobin} and {\em Azurin}-based junctions also displayed temperature-independent electron transport over a wide temperature range, reinforcing the exceptional conductive properties of these proteins\cite{raichlin2015protein,kayser2019solid}. In a single protein measurement, including a {\em Streptavidin} with biotin and an {\em Anti-Ebola Fab fragment}, conductance between electrodes attached to the proteins was in the order of nanosiemens over many nanometers, much higher than what could be explained by electron tunneling\cite{zhang2019role}. {\em OmcZ} nanowires, produced by Geobacter species, have high electron conductivity, exceeding 30 Siemens/cm. OmcZ is the only known nanowire-forming cytochrome essential for forming high-current-density biofilms, which require long-distance (\(\sim10 \mu m\)) extracellular electron transport. Its structure\cite{gu2023structure} reveals linear and closely stacked hemes that may account for its high conductivity. {We also included {\em Tubulin} and {\em Actin} in our investigation, as they have been suggested to be involved in biological quantum coherence \cite{HAMEROFF1996453, stuart1998quantum}, and {\em Insulin} as a reference.}

{Calculating the wave functions of proteins is challenging due to their large size, structural complexity, and the large number of electrons involved. Proteins contain thousands of atoms, resulting in a vast number of atomic orbitals and interactions that must be considered in quantum mechanical calculations. This complexity makes high-accuracy methods, such as Density Functional Theory (DFT) or post-Hartree-Fock methods, computationally prohibitive for entire proteins, as they require significant computational resources and time.} The semi-empirical extended Hückel method\cite{hoffmann1963extended} balances computational feasibility and insightful results, making it the best option for handling the complexity of proteins within a reasonable timeframe and with available computational resources. The extended Hückel method\cite{JPLowe} uses a linear combination of atomic orbitals (LCAO) approach. In this method, a molecular orbital \(\Phi_n\) with energy \(\varepsilon_n\) is expressed as a linear combination of atomic orbitals \(\chi_j\),
\[
\Phi_n=\sum_{j=1}^N C_{nj}\chi_j,
\]
where \(N\) is the number of atomic orbitals. These coefficients are determined by solving the Schrödinger equation for the molecule. The atomic orbital basis is typically the Slater-type orbitals. The extended Hückel method approximates the Hamiltonian matrix elements using 
\begin{equation}
    H_{ij}^{eH} = \begin{cases} 
H_{ii}^{eH} & \text{for } i = j, \\
K \cdot S_{ij} \cdot (H_{ii}^{eH} + H_{jj}^{eH})/2 & \text{for } i \ne j,
\end{cases}
\end{equation}
where the diagonal elements \(H_{ii}^{eH}\) are the ionization energies of the atomic orbitals\cite{JPLowe}, \(\mathbf{S}\) is the overlap matrix with elements \(S_{ij} = \int \chi_i(\mathbf{r}) \chi_j(\mathbf{r}) \, d\mathbf{r}\), and \(K\) is set to 1.75. The Hamiltonian \(\mathbf{H}^{eH}\) is on a non-orthogonal basis and should be transformed into an orthogonal basis using the Löwdin transformation:
\[
\chi_i' = \sum_{j=1}^N [S^{-1/2}]_{ij} \chi_j,
\]
where \(\mathbf{S}^{-1/2}\) is the inverse square root of the overlap matrix. The transformed Hamiltonian is:
\begin{equation}
    \mathbf{H} = \mathbf{S}^{-1/2} \mathbf{H}^{eH} \mathbf{S}^{-1/2}.
\end{equation}
Then the eigenenergies \(\varepsilon_n\) and normalized eigenvectors \(\varphi_i^n\) are solutions of the eigenproblem
\begin{equation}
    \varepsilon_n\varphi_i^n=\sum_{j=1}^N H_{ij}\varphi_j^n.
\end{equation}
The method requires the atomic coordinates of proteins as an input for computing the overlap and Hamiltonian matrices, which can be obtained, for example, from the Protein Data Bank (PDB)\cite{10.1093/nar/28.1.235}. {Numerical calculations were performed using the "Yet Another Extended Hückel Molecular Orbital Package" (YAeHMOP)\cite{yaehmop}.}

\begin{figure}[t]
\centering
\includegraphics[width=8cm]{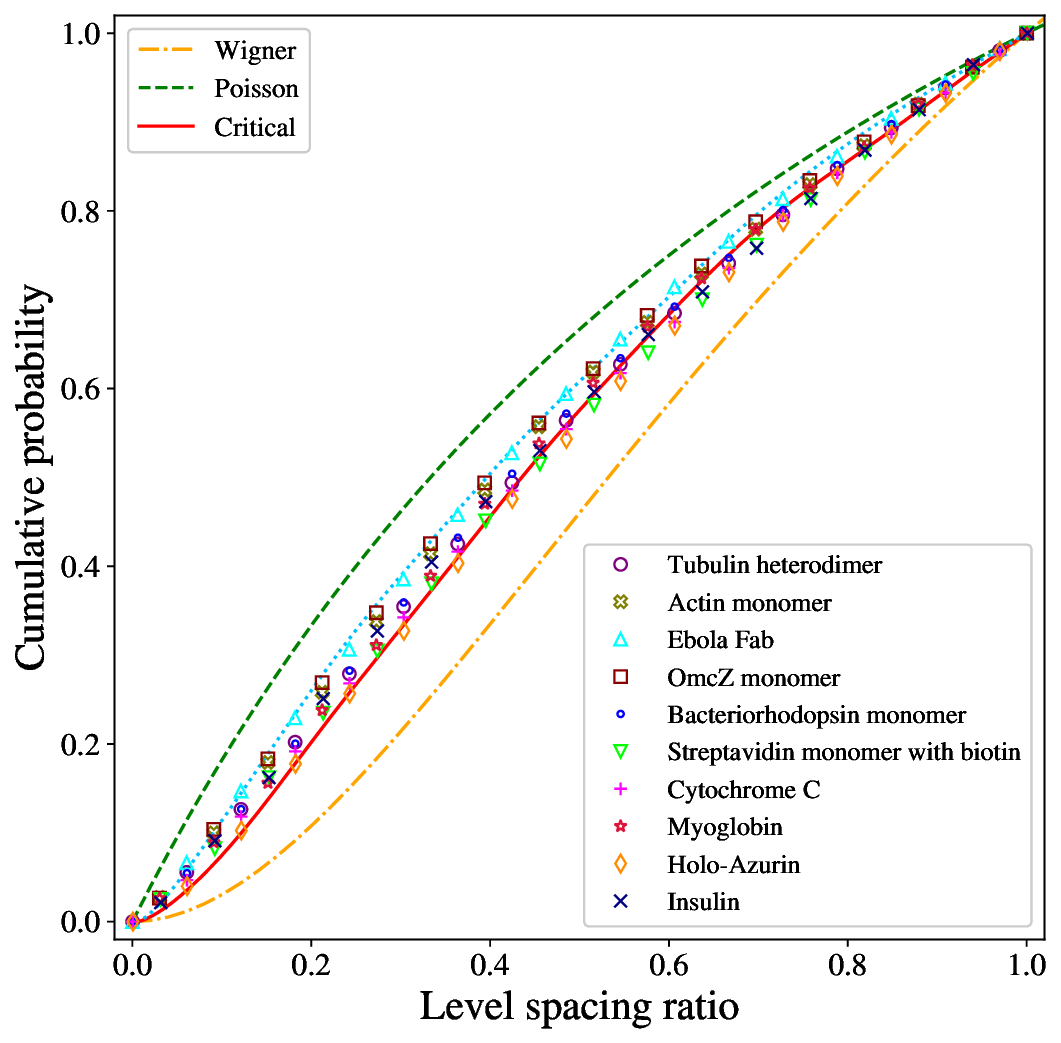}
\caption{Integrated level spacing ratio statistics for proteins in our investigation. The theoretical curves $I_W(r)$ for Wigner random matrix theory and for Poisson statistics $I_P(r)$ are shown. For the critical case, we used the critical value \(V_c\approx 16.4\) and a cube geometry with \(30\times 30 \times 30\) sites\cite{brandes1996critical} of the Anderson Hamiltonian (\ref{Anderson}). Out of the 27000 energy levels, we selected 5000 in the band center to represent \(I_c(r)\) as shown. We also show $I_V(r)$ corresponding to $V=17.9$ of the Anderson model. The experimentally determined atomic positions in proteins contain errors, causing a shift in the spectrum away from the critical point, which is generally more pronounced for larger proteins.}
\label{fig:ratio_stat}
\end{figure}

Energy level statistics is one of the key methods to characterize the Anderson transition. In the study of protein level statistics, analyzing the level spacing ratio has distinct advantages over the traditional spacing distribution approach. The level spacing ratio, defined as 
\[
r_i =\min\left\{\frac{s_{i+1}}{s_i},\frac{s_i}{s_{i+1}}\right\},
\]
where \(s_i = \varepsilon_{i+1} - \varepsilon_i\) represents the spacing between consecutive energy levels, provides a more robust and scale-invariant measure of level correlations. This method reduces the influence of local density variations and makes it easier to identify universal statistical properties. We introduce the integrated level spacing ratio statistics (ILSRS) as the probability that \(r_i<r\) or \(I(r)=\text{Prob}\{ r_i< r\}\). {In the insulating phase, the energy levels are randomly distributed and follow Poisson statistics. In this case, the ILSRS takes the simple form\cite{atas2013distribution}} 
\begin{equation}
    I_P(r)=\frac{2r}{1+r}.\label{Poisson}
\end{equation}
{In the metallic phase, the system can be characterized by random matrix Hamiltonians, whose level spacings follow the Gaussian Orthogonal Ensemble (GOE) predictions, which are well approximated by the Wigner surmise.}
The ILSRS can be well approximated\cite{atas2013distribution} by 
\begin{equation}
    I_W(r)=1+\frac{2r^3+3r^2-3r-2}{2(1+r+r^2)^{3/2}}.\label{GOE}
\end{equation}
In the case of the critical point of the Anderson MIT, there is no closed-form expression for the ILSRS. Therefore, we generated a 3D Anderson Hamiltonian\cite{shklovskii1993statistics}
\begin{equation}
    H=\sum_i V_i a_i^\dagger a_i-\sum_{\langle i,j\rangle}a_j^\dagger a_i,\label{Anderson}
\end{equation}
{ with nearest neighbor couplings and site energies uniformly distributed in the interval \(V_i\in [-V/2,+V/2]\). In the critical case $I_c(r)$ we used the crtitical value\cite{brandes1996critical} $V=V_c=16.4$. The ILSRS of proteins shown in Fig.~\ref{fig:ratio_stat} deviate from the Poisson and Wigner statistics, and all follow the critical Anderson curve.
Slight deviations from the critical $I_c(r)$ are observed for larger proteins due to the cumulative effect of the errors of experimentally determined atomic positions. To characterize the size of the deviation in terms of the Anderson model parameter, we also generated $I_V(r)$ using (\ref{Anderson}) and found the parameter $V=17.9$, which slightly deviates from the critical value towards the Poissonian and fits the data for the larger proteins.  }

\begin{figure}[htb]
\centering
\includegraphics[width=8cm]{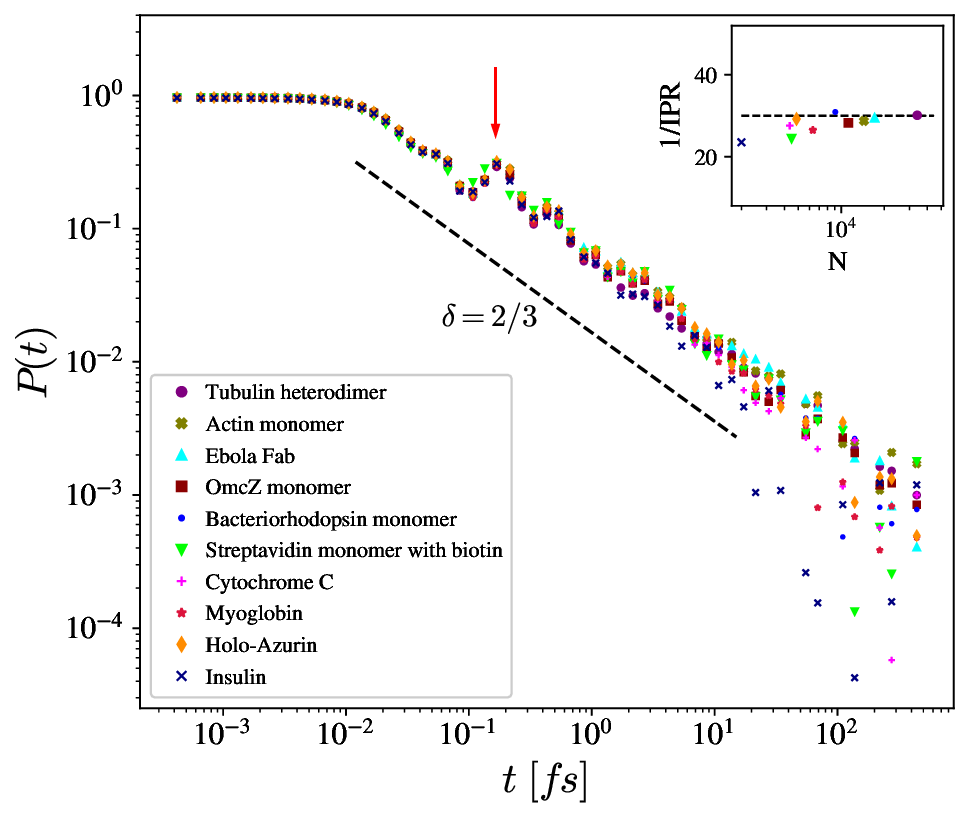}
\caption{The time-dependent part of the return probability for the eight proteins investigated. Notice that the curves for different proteins overlap up to about 10 fs. All curves show power law decay consistent with { \(\delta=0.66\pm0.03 \).} The initial part of the curve up to about 0.1685 fs corresponds to the electron diffusing within a single amino acid, where it peaks (red arrow), indicating a higher return probability due to the difficulty of leaving the initial amino acid. In the inset, the average localization length of the wave functions, i.e., \(1/I_2\), the inverse of the mean IPR of the proteins, is shown as a function of their number of atomic orbitals \(N\). The localization length is about 30 (dashed line), close to an amino acid's average number of atomic orbitals.}
\label{fig:return}
\end{figure}

Another key characteristic of the critical Anderson state is the anomalous diffusion of electrons due to the fractal nature of wave functions. The anomalous nature of diffusion manifests itself in the site-averaged return probability of electrons
\begin{equation}
P_R(t)=\frac{1}{N}\sum_i |K_{ii}(t)|^2,
\end{equation}
where 
\[
K_{ij}(t)=\sum_n\exp(-i\varepsilon_n t)\varphi_i^n\varphi^n_{j}
\]
is the propagator of the wave packet. This represents the likelihood that a wave packet remains at the point of its creation at \(t = 0\) after a time \(t\), or the probability that a random walker is located at its initial site at time \(t\). It can also be expressed as a constant plus a time-dependent part \(P_R(t)=I_2+P(t)\), where the constant part is the mean inverse participation ratio (IPR) of the wave functions  
\begin{equation}
    I_2=\frac{1}{N}\sum_{n=1}^N\sum_{i=1}^N |\varphi_i^n|^4,\label{IPR}
\end{equation}
and the time-dependent part    
\begin{equation}
P(t)=2\sum_{n>m} C_{n,m}\cos\left((\varepsilon_n-\varepsilon_m)t\right),\label{ft}
\end{equation}
is the Fourier transform of the density-density correlation function of the eigenfunctions
\begin{equation}
    C(\omega)=\sum_{n,m}C_{n,m}\delta\left(\omega-\varepsilon_n+\varepsilon_m\right),\label{denden}
\end{equation}
where
\[
C_{n,m}=\frac{1}{N}\sum_{i=1}^N|\varphi^n_i|^2|\varphi^m_i|^2.
\]
The decay of the return probability is characterized by the exponent \(\delta\)
\begin{equation}
    P(t)\sim t^{-\delta}.
\end{equation}
The exponent is \(\delta=d/2\) for extended states in the metallic phase, { where $d$ is the Euclidean dimension of the space} and \(\delta=d_2/d\) for critical states\cite{huckestein1994relation} characterized by the correlation fractal dimension \(d_2\) of the wave functions\cite{pook1991multifractality,pracz1996correlation}. In Fig.~\ref{fig:return}, we show the time-dependent part of the return probability \(P(t)\) for all proteins investigated. The initial part of the curve, up to 0.1685 fs, corresponds to the electron diffusion within a single amino acid. The bump around 0.1685 fs reflects the difficulty of leaving the initial amino acid and the higher probability of returning to the origin. The part beyond 0.1685 fs is related to large-scale inter-amino acid diffusion. Up to the cutoff time of about 10 fs, proteins of different sizes and shapes share the same universal curve. The exponent \(\delta\approx 2/3\) indicates that the fractal dimension of the wave functions is { \(d_2 = 2.0\pm 0.09\) (\(d=3\)),} approximately an integer.
    
\begin{figure}[bt]
\centering
\includegraphics[width=8cm]{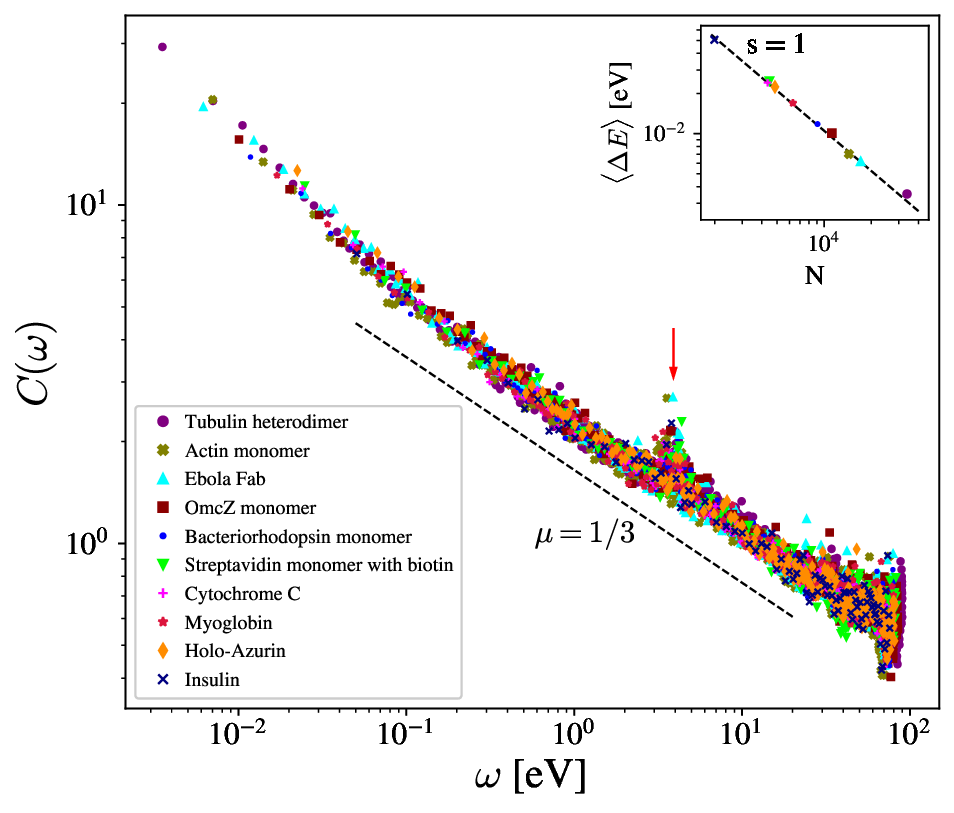}
\caption{Eigenfunction density-density correlation as a function of energy separation \(\omega\). \(\langle \Delta E\rangle\) is the mean level spacing. A peak (red arrow) indicates the point \(3.90\,\text{eV}=\hbar/(0.1685\,\text{fs})\) corresponding to the peak of the return probability. The distribution shows power-law scaling with exponent \(\mu\approx 1/3\) (dashed line). The inset shows the mean level spacing as a function of the protein size, consistent with \(\langle\Delta E\rangle \approx (105\,\text{eV})/N\) (dashed line).}
\label{fig:corr}
\end{figure}

The distribution of the density-density correlation of wave functions (\ref{denden}) further confirms these findings. Since the level spacing of proteins is highly variable throughout the energy spectrum, we study the wave function correlation between the \(l^{\mathrm{th}}\) neighbor energy levels 
\begin{equation}
    C(\omega)=\sum_{n=l+1}^{N} C_{n,n-l},
\end{equation}
where \(l=n-m\) and \(\omega=\langle\Delta E\rangle \cdot l\) is the mean distance between the energy levels \(\varepsilon_n\) and \(\varepsilon_m\) with \(\langle\Delta E\rangle=(\varepsilon_{\text{max}}-\varepsilon_{\text{min}})/N\)\cite{rem}. The correlation function is expected to show power-law scaling\cite{chalker1990scaling,kravtsov2010dynamical}
\begin{equation}
    C(\omega)\sim \omega^{-\mu},
\end{equation}
with exponent \(\mu=1-\delta\) since it is the Fourier transform (\ref{ft}) of the time-dependent part of the return probability. In Fig.~\ref{fig:corr}, the correlations for the investigated proteins display a universal power law with exponent \(\mu\approx 1/3\) up to a cutoff energy of 20 electronvolts. The mean level spacing scales as \(\langle\Delta E\rangle\sim N^{-1}\) following Weyl’s law\cite{roe1999}.

\begin{figure}[tb]
\centering
\includegraphics[width=8cm]{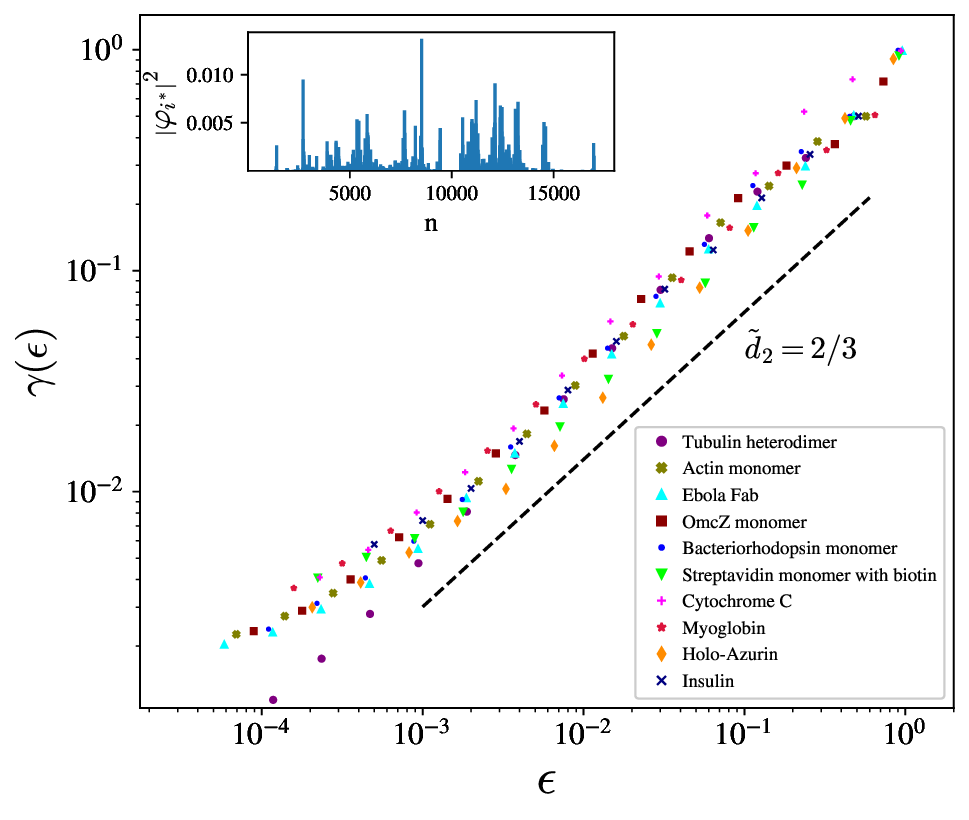}
\caption{Fractal dimension of the protein wave functions. The function \(\gamma(\epsilon)\) defined in (\ref{gamma}) is shown for all proteins investigated. The slope of the curve gives the dimension \(\tilde{d}_2=d_2/3\) which is consistent with \(\tilde{d}_2=2/3\) (dashed line). The inset shows the fractal wave function \(|\varphi_{i^*}^n|^2\) as a function of the index \(n\) for atomic orbital \(i^*=6234\) of the Ebola Fab protein.}
\label{fig:frac}
\end{figure}

Finally, we directly confirm the value of the correlation fractal dimension of the wave functions. {Measuring the fractal dimension in real space is complicated due to the intricate geometry of proteins.} Instead, we calculate the fractal dimension in energy space following Refs.\cite{ketzmerick1992slow,huckestein1994relation}. We select a site \(i^*\) for each protein for which the energy space inverse participation ratio is minimized, i.e. \(\min_i\sum_n |\varphi_i^n|^4\). Then, we group \(l\) neighboring energies into the same box. For each box, we calculate the probability 
\[
P_k=\sum_{n=n_k}^{n_{k+1}-1} |\varphi_{i^*}^n|^2,
\]
with \(n_k=(k-1)\cdot l+1\), and define \(\epsilon=l/N\) as the box size relative to the system size. The function
\begin{equation}
    \gamma(\epsilon)=\sum_{k=1}^{\lfloor N/l\rfloor}P_k^2\sim \epsilon^{\tilde{d}_2},\label{gamma}
\end{equation}
where \(\tilde{d}_2=d_2/3\) in 3D (\(d=3\))\cite{huckestein1994relation}, is then calculated. In Fig.~\ref{fig:frac}, the data are consistent with previous findings and confirm that \(d_2\approx 2\).

Remarkably, the fractal dimension is nearly an integer, unlike in the Anderson model (\ref{Anderson}), where \(d_2\approx 1.6\)\cite{parshin1999distribution} or in the quantum Hall system where \(d_2\approx 1.43\)\cite{pook1991multifractality}. For a typical fractal wave function of the protein in real space, integrating the probability of finding the electron within a ball of radius \(R\) scales as
\begin{equation}
  \int_{\text{ball}}d\mathbf{r}\,|\Phi_n(\mathbf{r})|^2\sim R^2,
\end{equation}
in contrast to the volume scaling \(\sim R^3\) for metallic conductors with extended wave functions. This surface scaling is a ``holographic'' property\cite{black} and likely reflects evolutionary optimization toward faster electron transport\cite{vattay2014quantum,vattay2015quantum}.


{ Findings presented in this paper demonstrate that a diverse range of proteins, crucial for biological electron transport, naturally reside in a quantum critical state analogous to the Anderson metal-insulator transition (MIT). This self-organization into criticality, without the need for external parameter tuning, is remarkable and carries profound implications for understanding their electron transport properties and, by extension, their biological functions. To appreciate these implications, it is valuable to consider the unique characteristics of electron transport at the critical point of the Anderson MIT. At the critical point, a system is delicately poised between metallic (diffusive) and insulating (localized) behavior. This state is not merely an intermediate phase but possesses distinct, universal transport characteristics. The observation that proteins exhibit hallmarks of this critical regime offers a new lens through which to view their electronic behavior. One of the defining features of the Anderson critical point is scale-invariant conductance. While perfect scale invariance to a universal critical conductance $g_c$ is an idealized limit, the tendency towards such behavior, as suggested by the proximity of protein level statistics to critical models (Fig. 1), implies a significant functional advantage. It suggests that the efficiency of electron transport through these proteins is inherently robust, less susceptible to minor variations in their length (within biological scales) or subtle conformational changes that are ubiquitous in dynamic biological environments. This intrinsic stability ensures reliable charge transfer, which is paramount for consistent biological processes like respiration, photosynthesis, and enzymatic catalysis. Furthermore, electron transport at criticality is characterized by anomalous diffusion, distinct from the ballistic motion in ideal conductors or the complete arrest in insulators. Our findings of a power-law decay in the site-averaged return probability, $P(t)\sim t^{-\delta}$ with $\delta\approx 2/3$ (Fig. 2), are a direct manifestation of this. This subdiffusive behavior indicates that while electrons are not freely diffusing as in a metal, they are also not strictly localized. This intermediate transport regime could be an evolutionary optimum for biological systems. It allows charges to traverse significant intramolecular distances efficiently, yet in a controlled manner, preventing overly rapid dissipation that might occur in a highly conductive metal, and avoiding the complete trapping of charge characteristic of an insulator. This ensures that electrons reach their intended destinations effectively without getting lost or moving too quickly for subsequent biochemical steps. The electronic wave functions at the critical point are also multifractal, possessing a complex, spatially fluctuating nature that is neither uniformly extended nor exponentially localized. Our calculation of the correlation fractal dimension $d_2\approx 2$ (Fig. 4) for these proteins is particularly striking. This integer value, suggesting a "holographic" area-law scaling for wave function probabilities ($ \int_{ball}dr|\Phi_{n}(r)|^{2}\sim R^{2} $), is unique compared to the non-integer $d_2$ values typically found in other 3D Anderson critical systems. This implies that electron transport in proteins occurs not just through a bulk material but is likely channeled along highly optimized, effectively lower-dimensional pathways embedded within the protein's three-dimensional structure. These multifractal pathways could minimize backscattering and ensure high transmission probability along specific routes, contributing to the observed long-range, temperature-independent conductance in many proteins. This "holographic" characteristic points towards an exceptional degree of evolutionary refinement for efficient electron channeling. Finally, the emergence of universal statistical properties (like level spacing ratios and aspects of conductance fluctuations) at criticality suggests that the observed transport characteristics might represent a fundamental design principle employed by nature across a variety of proteins involved in electron transfer. The fact that different proteins, despite their structural and functional diversity, converge to these critical signatures points to a common strategy for optimizing quantum transport. In essence, the positioning of proteins at a quantum critical point is not a mere physical curiosity but appears to be a sophisticated biological strategy. It endows them with a unique suite of transport properties—robustness, controlled yet efficient anomalous diffusion, and highly structured multifractal pathways—that are likely crucial for their biological efficacy. This perspective shifts our understanding of proteins from passive insulators or simple conductors to dynamic systems exquisitely tuned to leverage the subtle physics of quantum criticality. These insights not only deepen our comprehension of fundamental biological processes like energy conversion and signaling but also open exciting avenues for future research. The distinct "holographic" nature of criticality in proteins ($d_2\approx 2$) warrants further theoretical and experimental investigation to unravel its origins and full functional consequences. Furthermore, understanding how biological systems achieve and maintain this critical state could inspire the design of novel bioelectronic devices and nanostructured materials that harness quantum critical phenomena for enhanced efficiency, sensitivity, and tailored transport characteristics, potentially revolutionizing fields from neuromorphic computing to energy harvesting. The self-organized criticality observed in these life-sustaining molecules underscores a profound intersection of quantum physics and biology, promising a rich landscape for future discoveries.
}

\section*{Acknowledgments}
This research was supported by the Ministry of Culture and Innovation and the National Research, Development, and Innovation Office within the Quantum Information National Laboratory of Hungary (Grant No. 2022-2.1.1-NL-2022-00004).

While preparing this work, the authors used Grammarly (including its AI services) to improve grammar and text composition. After using this tool, the authors reviewed and edited the content as needed and took full responsibility for the publication's content. 


\begin{thebibliography}{100}
\bibitem{rem}
In practice, we calculate $\langle \Delta E\rangle$ by dropping the few highest energy levels of the proteins since those are statistical outliers due to the numerical instability of the high-lying antibonding molecular orbitals.
\bibitem{black}
We borrow this wording from black hole physics, where the Bekenstein-Hawking entropy is proportional to the area of the event horizon.
\end{thebibliography}

\end{document}